\documentclass[aps,pre,twocolumn,amsmath,showpacs]{revtex4}
\usepackage{graphicx}
\begin{document}

\title{Broadening of a nonequilibrium phase transition by extended structural defects}

\author{Thomas Vojta}
\affiliation{Department of Physics, University of Missouri-Rolla, Rolla, MO 65409}

\date{\today}

\begin{abstract}
We study the effects of quenched extended impurities on nonequilibrium phase transitions in the
directed percolation universality class. We show that these impurities have a dramatic effect:
they completely destroy the sharp phase transition by smearing. This is caused by rare strongly
coupled spatial regions which can undergo the phase transition independently from the bulk
system. We use extremal statistics to determine the stationary state as well as the dynamics in
the tail of the smeared transition, and we illustrate the results by computer simulations.
\end{abstract}

% insert suggested PACS numbers in braces on next line
% insert suggested keywords - APS authors don't need to do this
%\keywords{}

%\maketitle must follow title, authors, abstract, \pacs, and \keywords

\maketitle

% body of paper here - Use proper section commands
%%%%%%%%%%%%%%%%%%%%%%%%%%%%%%%%%%%%%%%%%%%%%%%%%%%%%%%%%%%%%%%%%%%%%%%%%%%%%%%%%%%%%%%%%

In nature, thermal equilibrium is more of an exception than the rule.
% While for many systems,equilibrium is a good approximation,
%other systems are far from equilibrium and show new and surprising behavior.
%
In recent years, phase transitions between different nonequilibrium states have become a topic
of great interest. A prominent class of nonequilibrium phase transitions separates active,
fluctuating states from inactive, absorbing states where fluctuations cease entirely. These
absorbing state transitions have applications ranging from physics to chemistry and to biology
\cite{chopard_book,marro_book,hinrichsen00,tauber}. The generic universality class for
absorbing state transitions is directed percolation (DP) \cite{dp}. According to a conjecture
by Janssen and Grassberger \cite{conjecture}, all absorbing state transitions with a scalar
order parameter, short-range interactions and no extra symmetries or conservation laws belong
to this class. Examples include the transitions in the contact process \cite{contact},
catalytic reactions \cite{ziff}, interface growth \cite{tang}, or turbulence \cite{turb}.
However, despite its ubiquity in theory and simulations, clearcut experimental realizations of
the DP universality class are strangely lacking \cite{hinrichsen_exp}. The only verification so
far seems to be found in the spatio-temporal intermittency in ferrofluidic spikes
\cite{spikes}.

A possible reason for this apparent discrepancy are impurities, i.e., quenched spatial
disorder. According to the Harris criterion \cite{harris,noest}, the DP universality class is
unstable against disorder, because the (spatial) correlation length exponent $\nu_\perp$
violates the inequality $\nu_\perp > 2/d$ for all spatial dimensionalities $d<4$. Indeed, in
the corresponding field theory, spatial disorder leads to runaway flow of the renormalization
group (RG) equations \cite{janssen97}, destroying the DP behavior. Several other studies
\cite{bramson,moreira,webman,cafiero} agreed on the instability of DP against spatial disorder,
but a consistent picture has been slow to evolve. Recently, Hooyberghs {\it et al.} applied the
Hamiltonian formalism \cite{alcaraz} to the contact process with spatial disorder
\cite{hooyberghs}. Using the Ma-Dasgupta-Hu strong-disorder RG \cite{SDRG} these authors showed
that the transition (at least for sufficiently strong disorder) is controlled by an exotic
infinite-randomness fixed point with activated rather than the usual power-law scaling.
In many real systems, the disorder does not consist of point defects but of dislocations,
disordered layers, or grain boundaries. The effects of such extended defects are generically
stronger than that of uncorrelated disorder, as has been shown by detailed studies of
equilibrium systems ranging from the exactly solved McCoy-Wu model \cite{McCoyWu} and several
RG studies \cite{Lubensky,Dorogovtsev,BoyCardy,DeCesare} to the discovery of
infinite-randomness critical behavior in the corresponding quantum Ising model \cite{dsf9295}.

In this paper, we show that extended defects have an even more dramatic effect on
non-equilibrium phase transitions in the DP universality class;  they destroy the sharp
transition by smearing. This is caused by phenomena similar to but stronger than the usual
Griffiths effects \cite{Griffiths,noest}: rare strongly coupled spatial regions can undergo the
transition independently from the bulk system. In the tail of the smeared transition, the
spatial density distribution is very inhomogeneous, with the average stationary density and the
survival probability depending exponentially on the control parameter. The approach of the
average density to this exponentially small stationary value occurs in two stages, a stretched
exponential decay at intermediate times, followed by power-law behavior at late times. In the
following, we derive these results for a disordered contact process, illustrate them by
computer simulations, and discuss their generality and importance.

Our starting point is the clean contact process \cite{contact}, a prototypical system in the DP
universality class. It is defined on a $d$-dimensional hypercubic lattice. Each site $\mathbf
r$ can be vacant or active, i.e, occupied by a particle. During the time evolution, particles
are created at vacant sites at a rate $\lambda n/ (2d)$ where $n$ is the number of active
nearest neighbor sites and the `birth rate' $\lambda$ is the control parameter. Particles are
annihilated at unit rate. For small $\lambda$, annihilation dominates, and the absorbing state
without any particles is the only steady state. For large $\lambda$ there is a steady state
with finite particle density (active phase). Both phases are separated by a nonequilibrium
phase transition in the DP universality class at $\lambda=\lambda_c^0$.

We introduce quenched spatial disorder by making the birth rate $\lambda$ a random function of
the lattice site. Extended impurities can be described by disorder perfectly correlated in
$d_c$ dimensions, but uncorrelated in the remaining $d_r=d-d_c$ dimensions. $\lambda$ is thus a
function of ${\mathbf r}_r$ which is the projection of the position vector $\mathbf r$ on the
uncorrelated directions. For definiteness, we assume that the $\lambda({\mathbf r}_r)$ have a
binary probability distribution
\begin{equation}
P[\lambda({\mathbf r}_r)] = (1-p)\, \delta[\lambda({\mathbf r}_r)-\lambda] + p\,
\delta[\lambda({\mathbf r}_r) - c\lambda]
\end{equation}
where $p$ and $c$ are constants between 0 and 1. In other words, there are extended impurities
of density $p$ where the birth rate $\lambda$ is reduced by a factor $c$.

Let us now consider the effects of rare disorder fluctuations in this system. In analogy to the
Griffiths phenomena \cite{Griffiths,noest}, there is a small but finite probability for finding
large spatial regions devoid of impurities. These rare regions can be locally in the active
phase, even if the bulk system is still in the inactive phase. For the largest rare regions
this starts to happen when $\lambda$ crosses the clean critical point $\lambda_c^0$. Since the
impurities in our system are extended, each rare region is infinite in $d_c$ dimensions but
finite in the remaining $d_r$ dimensions. This is a crucial difference to systems with
uncorrelated disorder, where the rare regions are finite. In our system, each rare region can
therefore undergo a real phase transition {\em independently} of the rest of the system. Thus,
those rare regions that are locally in the ordered phase will have a true nonzero stationary
density, even if the bulk system is still in the inactive phase.

The resulting global phase transition is very different from a conventional continuous phase
transition, where a nonzero order parameter develops as a collective effect of the entire
system, accompanied by a diverging correlation length in all directions. In contrast, in our
system, the order parameter develops very inhomogeneously in space with different parts of the
system (i.e., different ${\bf r}_r$ regions) ordering independently at different $\lambda$.
Correspondingly, the correlation length in the uncorrelated directions remains finite across
the transition. This defines a smeared transition. Thus, extended impurities lead to a smearing
of the DP phase transition.

We now use extremal statistics to derive the  properties in the tail of the smeared transition,
i.e., in the parameter region where a few active rare regions exist, but their density is so
small that they can be treated as independent. We start with the stationary behavior. The
probability $w$ for finding a rare region of linear size $L_r$ devoid of impurities is, up to
pre-exponential factors, given by
\begin{equation}
w \sim \exp(-\tilde p L_r^{d_r}) \label{eq:rr}
\end{equation}
with  $\tilde p = -\ln(1-p)$. As discussed above, such a region undergoes a true phase
transition to the active phase at some $\lambda_c(L_r)>\lambda_c^0$. According to finite-size
scaling \cite{barber},
\begin{equation}
\lambda_c(L_r) - \lambda_c^0 = A L_r^{-\phi}~, \label{eq:shift}
\end{equation}
where $\phi$ is the clean ($d$-dimensional) finite-size scaling shift exponent and $A$ is the
amplitude for the crossover from a $d$-dimensional bulk system to a `slab' infinite in $d_c$
dimensions but but finite in $d_r$ dimensions. If the total dimensionality $d=d_c+d_r<4$,
hyperscaling is valid and $\phi = 1/\nu_\bot$ which we assume from now on. Combining
(\ref{eq:rr}) and (\ref{eq:shift}) we obtain the probability for finding a rare region which
becomes active at $\lambda_c$ as
\begin{equation}
w(\lambda_c) \sim \exp \left( -B (\lambda_c -\lambda_c^0)^{-d_r\nu_\bot} \right)
\label{eq:w_lam}
\end{equation}
for $\lambda_c-\lambda_c^0 \to 0+$. Here $B= \tilde p A^{d_r\nu_\bot}$. The total (average)
density $\rho$ at a certain $\lambda$ is obtained by summing over all active rare regions,
i.e., all regions with $\lambda_c<\lambda$. Since the functional dependence on $\lambda$ of the
density on any given active island is of power-law type it does not enter the leading
exponentials but only the pre-exponential factors. Thus, the stationary density develops an
exponential tail,
\begin{equation}
\rho(\lambda) \sim \exp \left( -B (\lambda-\lambda_c^0)^{-d_r\nu_{\bot}} \right)~,
\label{eq:rho}
\end{equation}
reaching to the clean critical point $\lambda_c^0$. Analogous arguments can be made for the
survival probability $P(\lambda$) of a single seed site. If the seed site is on an active rare
region it will survive with a probability that depends on $\lambda$ with a power law. If is not
on an active rare region, the seed will die. To exponential accuracy the survival probability
is thus also given by (\ref{eq:rho}).

The local (spatial) density distribution in the tail of the smeared transition is very
inhomogeneous. On active rare regions, the density is of the same order of magnitude as in the
clean system. Away from these regions it decays exponentially. The typical local density
$\rho_{\rm typ}$ can be estimated from the typical distance of any point from the nearest
active rare region. From (\ref{eq:w_lam}) we obtain
\begin{equation}
r_{\rm typ} \sim \exp \left[ B  (\lambda-\lambda_c^0)^{-d_r\nu_\bot}/{d_r} \right]~.
\end{equation}
At this distance, the local density has decayed to
\begin{equation}
\rho_{\rm typ} \sim e^{-r_{\rm typ}/\xi_0} \sim \exp \left\{ -C \exp \left[ B
(\lambda-\lambda_c^0)^{-d_r\nu_\bot} /{d_r} \right]\right\}
\end{equation}
where $\xi_0$ is the bulk correlation length (which is finite and changes slowly across the
smeared transition) and $C$ is a constant. A comparison with (\ref{eq:rho}) shows that the
relation between the typical and the average density is exponential, $|\log \rho_{\rm typ}|
\sim \rho^{-1/d_r}$, indicating an extremely broad local density distribution.

\begin{figure*}[t]
\includegraphics[height=4.7cm]{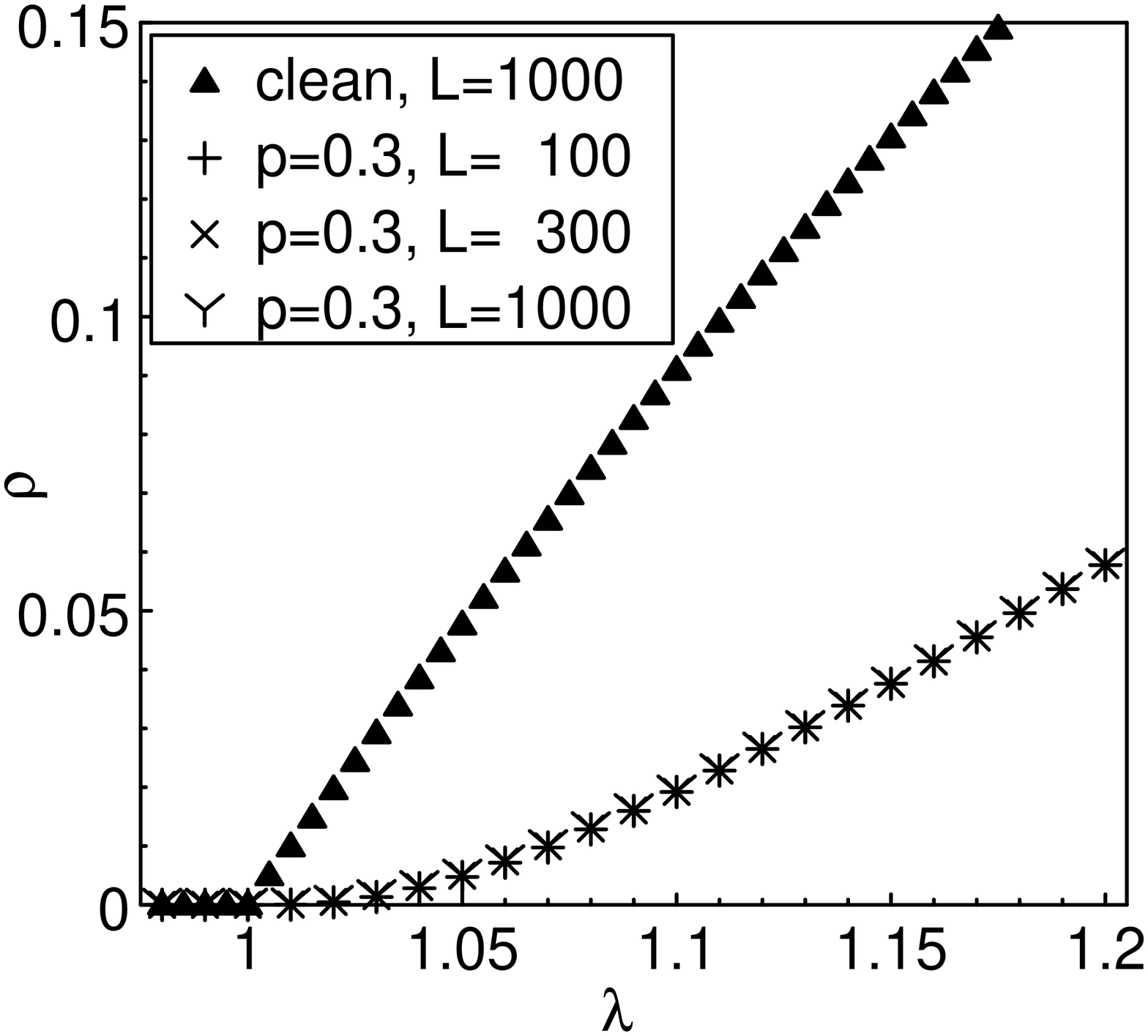}
\includegraphics[height=4.7cm]{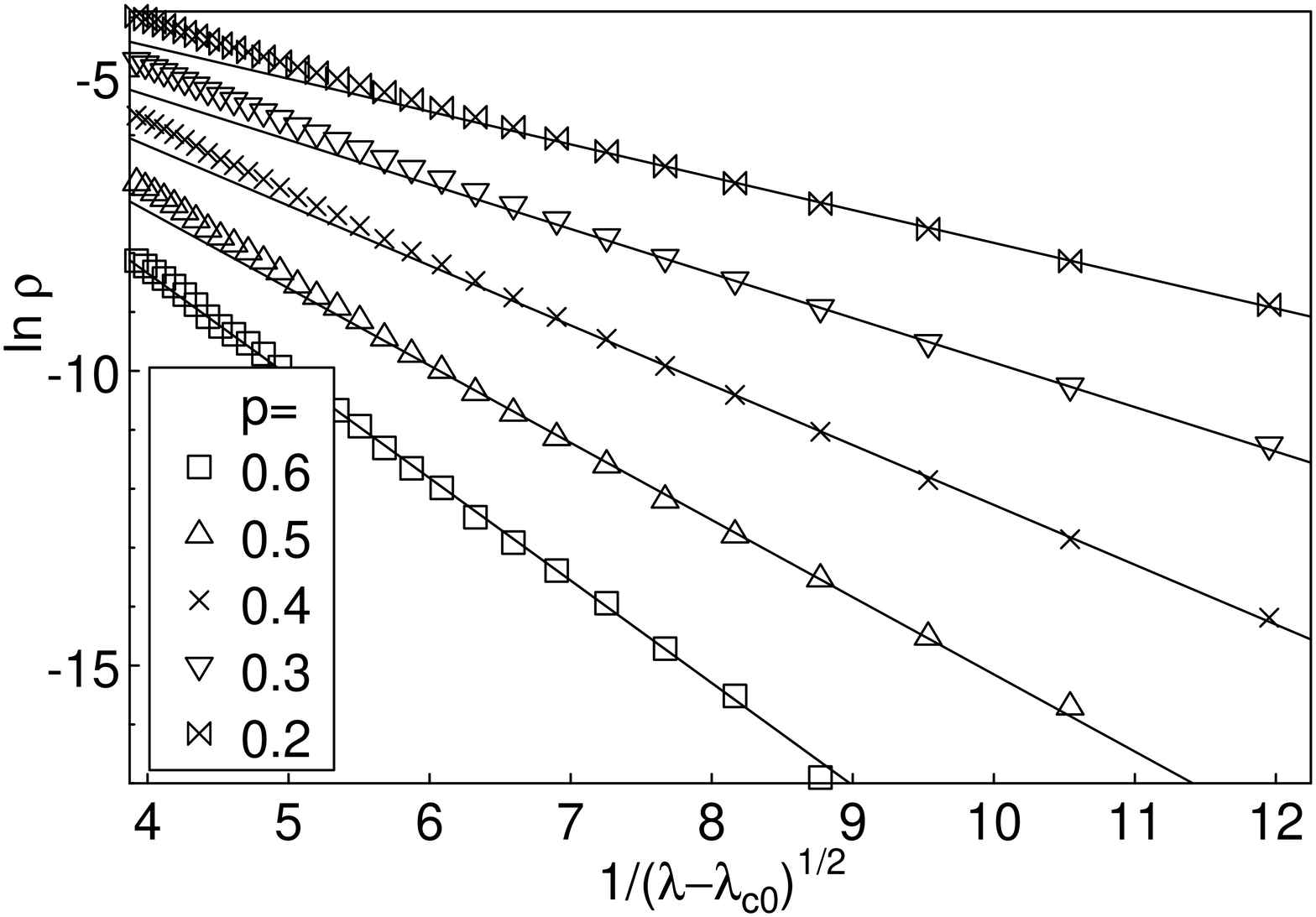}
\includegraphics[height=4.7cm]{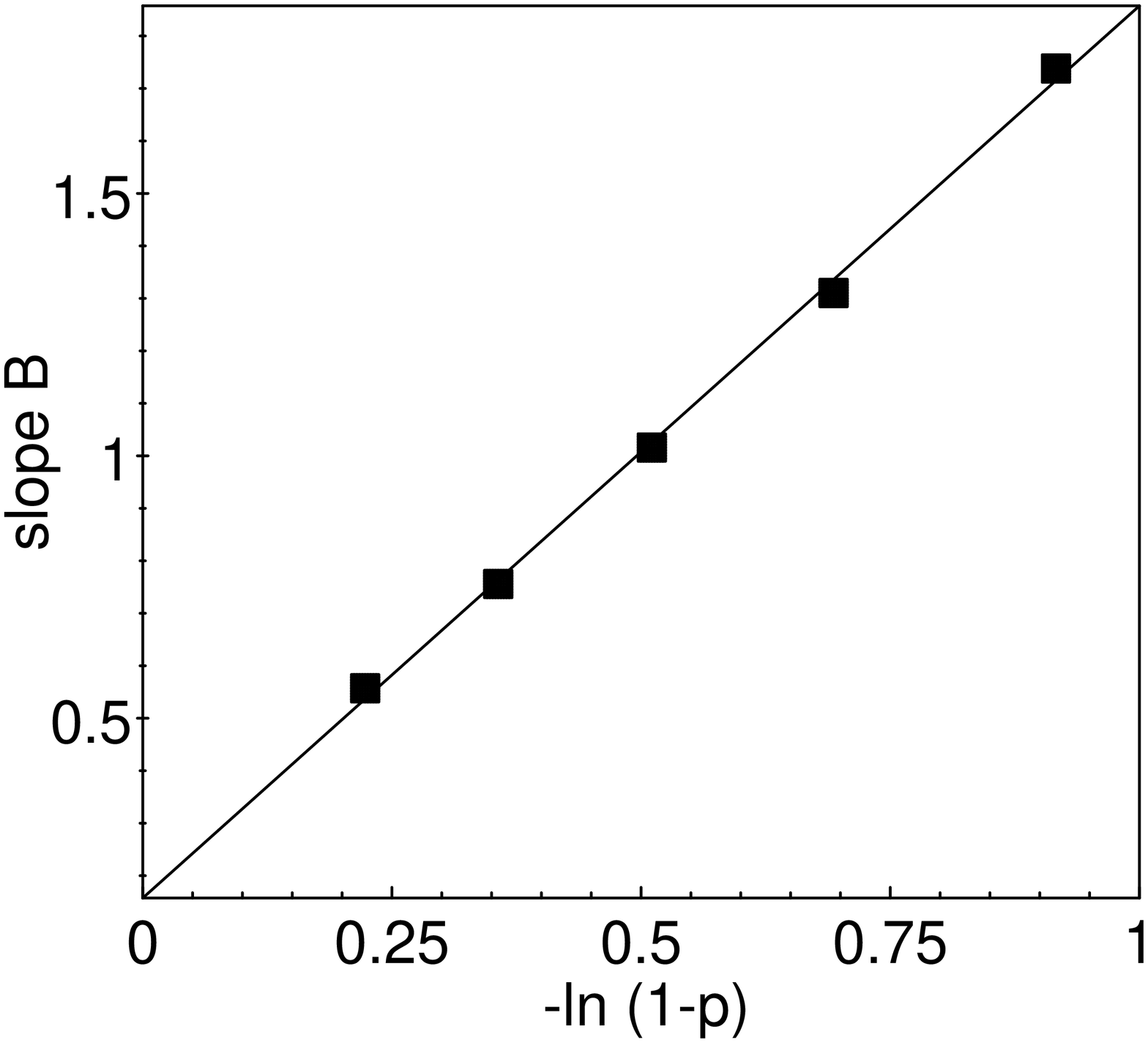}
\caption{Left: Overview of the steady state density of a clean ($p=0$) and a diluted ($p=0.3$)
system.  Center: Logarithm of the density as a function of $(\lambda-\lambda_c^0)^{-1/2}$ for
several dilutions $p$ and $L=10^4$. The data are averages over 100 disorder realizations. The
solid lines are fits to (\ref{eq:rho}) with $d_r\nu_\bot=1/2$. Right: Decay constant $B$ as a
function of $-\ln(1-p)$.} \label{fig:stationary}
\end{figure*}

We now turn to the dynamics in the tail of the smeared transition. The long-time decay of the
density is dominated by the rare regions while the bulk contribution decays exponentially.
According to finite size scaling \cite{barber}, the behavior of the correlation time $\xi_t$ of
a single rare region of size $L_r$ in the vicinity of the clean bulk critical point can be
modelled by
\begin{equation}
\xi_t(\Delta, L_r) \sim L_r^{(z\nu_\bot - \tilde z\tilde \nu_\bot)/\nu_\bot}
   \left|  \Delta - A L_r ^{-1/\nu_\bot}\right|^{-\tilde z \tilde \nu_\bot}~. \label{eq:xit}
\end{equation}
Here $\Delta=\lambda-\lambda_c^0>0$, $z$ is the $d$-dimensional bulk dynamical critical
exponent, and $\tilde \nu_\bot$ and $\tilde z$ are the correlation length and dynamical
exponents of a $d_r$-dimensional system. Let us first consider the time evolution of the
density at $\lambda=\lambda_c^0$. For $\Delta=0$, the correlation time (\ref{eq:xit})
simplifies to $\xi_t \sim L_r^z$. To exponential accuracy, the time dependence of the average
density is obtained by
\begin{equation}
\rho(t) \sim \int dL_r ~\exp \left (-\tilde p L_r^{d_r} -D t/L_r^z \right)
\end{equation}
where $D$ is a constant. Using the saddle point method to evaluate this integral, we find the
leading long-time decay of the density to be given by a stretched exponential,
\begin{equation}
\ln \rho(t) \sim - t^{d_r/(d_r+z)}~. \label{eq:stretched}
\end{equation}

For $\lambda>\lambda_c^0$, we repeat the saddle point analysis with the full expression
(\ref{eq:xit}) for the correlation length. For intermediate times $t<t_x\sim
(\lambda-\lambda_c^0)^{-(d_r+z)\nu_\bot}$ the decay of the average density is still given by
the stretched exponential (\ref{eq:stretched}). For times larger than the crossover time $t_x$
the system realizes that some of the rare regions are in the active phase and contribute to a
finite steady state density. The approach of the average density to this steady state value is
characterized by a power-law.
\begin{equation}
\rho(t) - \rho(\infty) \sim t^{-\psi}~. \label{eq:power}
\end{equation}
The value of $\psi$ cannot be found by our methods since it depends on the neglected
pre-exponential factors.

We now illustrate the smearing of the phase transition by results of a computer simulation of a
2d contact process with linear defects ($d_c=d_r=1$). To reach the rather large system sizes
necessary to observe exponentially rare events, we consider a version of the contact process
with infinite-range couplings in the correlated direction (parallel to the impurities) but
nearest-neighbor couplings in the uncorrelated direction (perpendicular to the defect lines).
While this infinite-range model will not be quantitatively comparable to a short-range contact
process, it provides a simple example for the smearing mechanism. Moreover, since the smearing
only relies on a single rare region undergoing a true phase transition, we expect that the
results will qualitatively valid for a short-range contact process, too (with the appropriate
changes to the exponents).

Because the couplings in the correlated direction are of infinite range, this dimension can be
treated exactly in mean-field theory. This leads to a set of coupled local mean-field equations
for the local densities $\rho_x$,
\begin{equation}
\frac \partial {\partial t} \rho_x = - \rho_x +\frac {\lambda(x)} 4 \,(1-\rho_x)\,
   \left( \rho_{x-1} + 2\rho_x +\rho_{x+1} \right)~. \label{eq:mf}
\end{equation}
These equations can be easily solved numerically. We study systems with several dilutions
$p=0.2\ldots 0.6$ and sizes of up to $L=10^6$ in the uncorrelated direction; the impurity
strength is $c=0.2$ for all calculations.

To determine the stationary state we solve the equations $\frac
\partial {\partial t} \rho_x =0 $ in a self-consistency cycle. Our results are summarized in
Fig.\ \ref{fig:stationary}. The left panel shows the total (average) density $\rho$ for a clean
($p=0$) and a dirty ($p=0.3$) system. The clean system has the expected sharp phase transition
at $\lambda=\lambda_c^0=1$ with the mean-field critical exponent $\beta=1$. The data of the
disordered system seem to suggest a transition at $\lambda \approx 1.04$. However, a closer
inspection shows that the singularity is smeared. Note that the density data are essentially
size-independent. Therefore, the observed rounding cannot be due  to finite-size effects. We
conclude that the smearing is an intrinsic effect of the infinite system. For comparison with
the analytical prediction (\ref{eq:rho}), the central panel shows the logarithm of the total
density as a function of $(\lambda-\lambda_c^0)^{-1/2}$ for different impurity concentrations
$p$. Note that in our infinite-range model $\nu_\bot=1/\phi=1/2$. The data follow eq.\
(\ref{eq:rho}) over several orders of magnitude in $\rho$. Fits of the data to (\ref{eq:rho})
are used to determine the decay constants $B$. The right panel of Fig.\ \ref{fig:stationary}
shows that these decay constants depend linearly on $\tilde p =- \ln(1-p)$, as predicted.

\begin{figure}
\includegraphics[height=4.7cm]{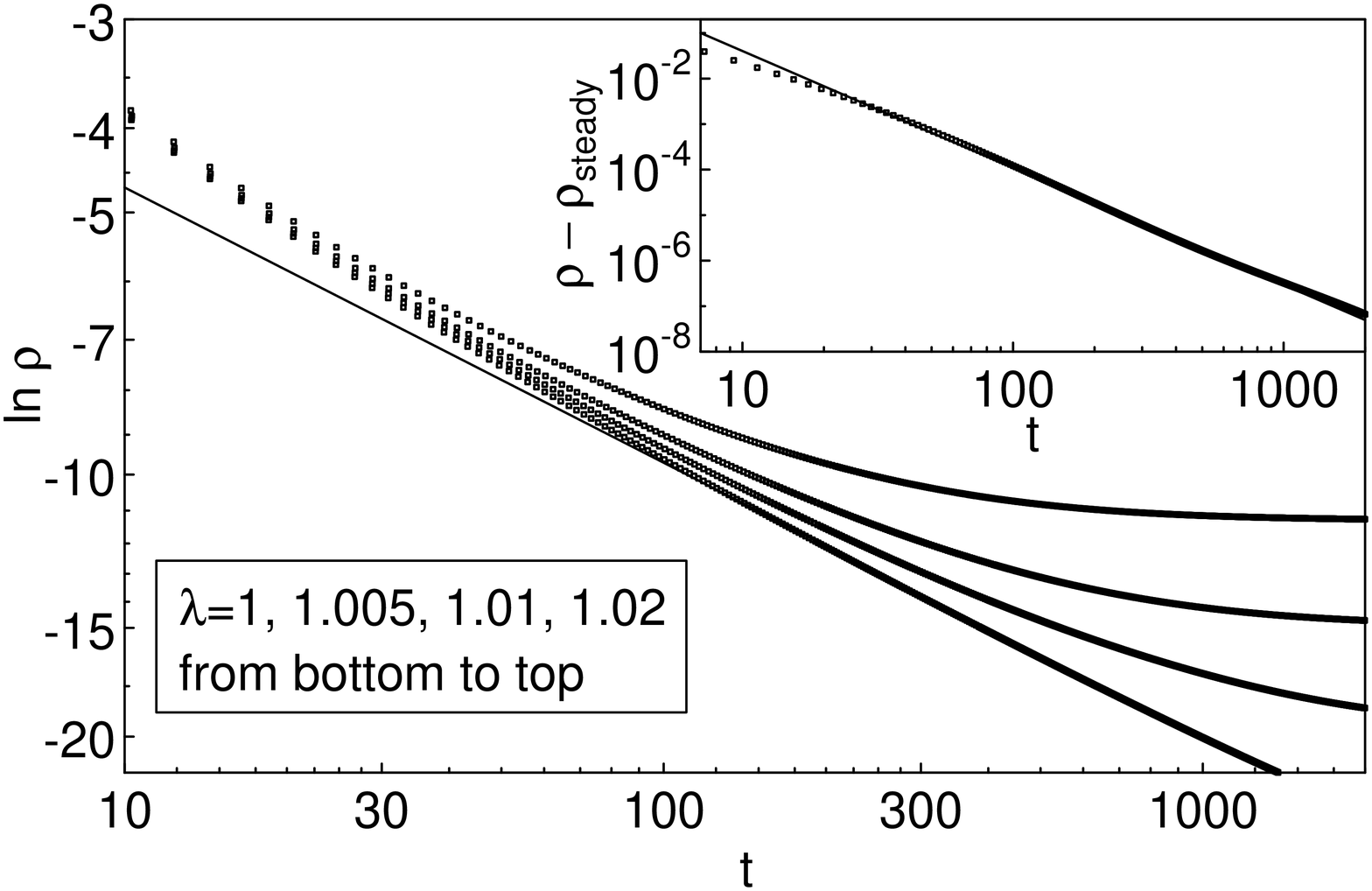}
\caption{Density $\rho$ vs.\ time $t$ for a system of size $L=10^6$, dilution $p=0.5$ and
several $\lambda$ (averages over 25 disorder realizations). Solid line: Fit of the $\lambda=1$
data ($t>100$) to eq.  (\ref{eq:stretched}) giving an exponent of approx. 0.32. Inset: Approach
to the steady state density for $\lambda=1.01$. Solid line: Fit of the data for $t>100$ to
(\ref{eq:power}), giving an exponent of $\psi \approx 2.6$.} \label{fig:evo}
\end{figure}

To study the time evolution we numerically integrate the local mean-field equations
(\ref{eq:mf}), starting from a homogeneous initial state with $\rho=1$. Fig.\ \ref{fig:evo}
summarizes our results for a system of size $L=10^6$ with dilution $p=0.5$.
The main panel shows a log-log plot of $\ln \rho$ vs. $t$. This allows us to test the stretched
exponential behavior  predicted  in (\ref{eq:stretched}) for the time dependence of $\rho$ at
the clean critical coupling $\lambda=\lambda_c^0=1$. We find that the data indeed follow a
stretched exponential with an exponent of approximately 0.32 in excellent agreement with the
analytical prediction $d_r/(d_r+z)=1/3$. For $\lambda>\lambda_c^0$, the decay initially follows
the stretched exponential, but eventually the density approaches its finite steady state value.
The inset of  Fig.\ \ref{fig:evo} shows a log-log plot of $\rho(t)-\rho_{\rm steady}$ vs. $t$.
The data clearly display power-law behavior in agreement with (\ref{eq:power}). A fit to this
equation gives an exponent value of $\psi \approx 2.6$.

To summarize, we have demonstrated that extended impurities destroy the sharp DP phase
transition in the contact process by smearing. In the remaining paragraphs we discuss the
generality of our findings as well as their relation to Griffiths effects
\cite{Griffiths,noest}.
The origins of Griffiths effects and the smearing found here are very similar, both are caused
by rare large spatial regions which are locally in the ordered phase. The difference between
them is a result of disorder correlations. For uncorrelated disorder, the rare regions are of
finite size and cannot undergo a true phase transition. Instead, they fluctuate slowly. In
contrast, if the rare regions are infinite in at least one dimension, a stronger effect occurs:
each rare region can independently develop a nonzero steady state density. This leads to a
smearing of the global transition.

The smearing mechanism found here relies only on the existence of a true phase transition on an
isolated rare region. It should therefore apply not only to the DP universality class, but to
an entire family of nonequilibrium universality classes for spreading processes and
reaction-diffusion systems. Note that while the presence or absence of smearing is universal in
the sense of critical phenomena (it depends on symmetries and dimensionality only), the
functional form of the density and other observables is {\em not} universal, it depends on the
details of the disorder distribution \cite{us_unpublished}. Smearing phenomena similar to the
one found here can also occur at equilibrium phase transitions. At quantum phase transitions in
itinerant electron systems, even point-like impurities can lead to smearing \cite{us_rounding}.
In contrast, for the classical Ising (Heisenberg) universality class, the impurities have to be
at least 2d (3d) for the transition to be smeared which makes the phenomenon less likely to be
observed \cite{us_planar}.

In conclusion, extended defects destroy the DP transition by smearing and lead to a
(nonuniversal) exponential dependence of the density and other quantities on the control
parameter. We suggest that this disorder-induced smearing may be related to the striking
absence of DP scaling \cite{hinrichsen_exp} in at least some of the experiments.

 We acknowledge stimulating discussions with U. T\"auber. This work was supported in
part by the University of Missouri Research Board.
%
%
% Create the reference section using BibTeX:

\end{document}